%% file: elspaper.tex
\documentclass[preprint,12pt]{elsarticle}

\usepackage{kurz}
\usepackage{amssymb}
\usepackage{mathrsfs}
\usepackage{multirow}    % BY THOMAS
\usepackage{stmaryrd}
\usepackage{amsmath}
\usepackage{caption}  
\usepackage{a4wide}
\biboptions{numbers,sort&compress}
\usepackage{upgreek}
\usepackage{longtable}

\journal{arXiv (as a draft)}
\begin{document}
\begin{frontmatter}



\title{A softening-healing law for self-healing quasi-brittle materials: analyzing with Strong Discontinuity embedded Approach}


\author[Leibniz]{Yiming Zhang \corref{cor}}
\author[Leibniz,tj,tj2]{Xiaoying Zhuang \corref{cor}}

\cortext[cor]{Corresponding authors:\\ \mbox{Yiming Zhang}, \mbox{+49 511 762-17535}, \mbox{Yiming.Zhang@ikm.uni-hannover.de};\\ \mbox{Xiaoying Zhuang}, \mbox{+49 511 762-19589}, \mbox{Xiaoying.Zhuang@gmail.com}}

\address[Leibniz]{Institute of Continuum Mechanics, Leibniz Universit\"{a}t Hannover, Appelstra{\ss}e 11, 30157~Hannover,~Germany}
\address[tj]{Department of Geotechnical Engineering, Tongji University, Siping Road 1239, 200092~Shanghai,~P.R.China}
\address[tj2]{Key Laboratory of Geotechnical and Underground Engineering of Ministry of Education, Tongji University, 200092~Shanghai, ~P.R.China}

\begin{abstract}
\input{abstract}

\end{abstract}

%All numerical models focus on answering following question i) What is the main driving force of spalling.  ii) When will the spalling happen under specific loading.  iii) What kind of protection method will alleviate spalling.

\end{frontmatter}
\clearpage

\setcounter{footnote}{0}
\input{content}

\clearpage
\appendix


%\bibliographystyle{elsarticle/model3a-num-names}
\clearpage
\bibliographystyle{ieeetr}
\bibliography{Reference}






% \end{thebibliography}

\end{document}

%% file: abstract.tex
Quasi-brittle materials such as concrete suffer from cracks during their life cycle, requiring great cost for conventional maintenance or replacement.  In the last decades, self-healing materials are developed which are capable of filling and healing the cracks and regaining part of the stiffness and strength automatically after getting damaged, bringing the possibility of maintenance-free materials and structures.  

In this paper, a time dependent softening-healing law for self-healing quasi-brittle materials is presented by introducing limited material parameters with clear physical background.  Strong Discontinuity embedded Approach (SDA) is adopted for evaluating the reliability of the model.  In the numerical studies, values of healing parameters are firstly obtained by back analysis of experimental results of self-healing beams.  Then numerical models regarding concrete members and structures built with self-healing and non-healing materials are simulated and compared for showing the capability of the self-healing material.

%% file: content.tex
\section{Introduction}
In engineering practices, the damage of quasi-brittle materials such as concrete, glass and ceramic manifests itself in the form of micro and macro cracks, jeopardizing the serviceability and durability of the structures and resulting into considerable maintenance and repair costs \cite{Ahn:01}.  Self-healing quasi-brittle materials attracted great interests in these years, which are capable of filling and repairing the cracks automatically, making it possible to greatly save the life term costs even build maintenance-free structures \cite{Breugel:01}.

The author of \cite{Dry:01} produced the first self-healing cement in 1990s.  After then, different self-healing strategies for quasi-brittle materials, especially for cementitious materials are presented, mainly including
\begin{itemize}
	\item 
	implementations of hollow fibers or micro capsules carrying bonding medium such as resin or component promoting chemical reaction of original material such as water for high strength concrete \cite{WhiteS:01,Tittelboom:01,Pang:01,HuangHaoliang:01,HuangHaoliang:02};
	\item
	utilizing bacteria for promoting the precipitation of calcium carbonate, which are commonly carried by micro capsules \cite{WangJianyun:01,Tziviloglou:01,WangJianyun:02,WangJianyun:03,Jonkers:01};
	\item
	using shape memory materials for providing traction between and closing the cracks \cite{SongG:01,Jefferson:01,Isaacs:01,Dunn:01};
\end{itemize}
with detailed discussions and comparisons provided in \cite{Tittelboom:02,Tang:01,WuMin:02,HuangHaoliang:04,Tittelboom:03,Schlangen:02}.  In these strategies, there are not enough results about the mechanical recovery regarding the bacteria healing method, although it has been proved to be very effective for controlling the long term mass transport e.g. permeation and diffusion inside crack opening.  On the other hand, the other two methods refer to conspicuous recovery of mechanical properties of the damage material as shown in experimental investigations.  

For numerically simulating the damage-healing behavior, based on Continuum Damage Mechanics (CDM), a consistent thermodynamic framework coined Continuum Damage-Healing Mechanics (CDHM) was firstly presented in \cite{Barbero:01}, which considered healing as a kind of ``negative damage" and presented a general framework for describing the degradation-healing of composites.  Belong to the family of CDHM, more recent methods are presented in these years \cite{Voyiadjis:01,Voyiadjis:02,Voyiadjis:03,Darabi:01,XuWei:01} including a version coined Cohesive Zone Damage-Healing model \cite{Alsheghri:01,Alsheghri:02}.  Based on CDHM, in \cite{Hazelwood:01} Fracture Process Zone (FPZ) elements and Shape Memory Polymer (SMP) tendon elements are respectively built and prescribed at the damage domain of the self-healing beam, for predicting the long-term mechanical behaviour of SMP reinforced concrete beam, as a recent engineering application.  CDHM was built based on CDM and inherits the basic assumptions of the latter, which introduces internal state variables and considers the damage-healing as reduction-increasing of stiffness and strength.  And the corresponding constitutive relationship is commonly built for strain and stress.

During damage of quasi-brittle materials, micro-defects merge into macro-cracks and the size of fracture zone reduces from finite width to almost zero in a very limited loading period, the behavior of which is localized and anisotropic, highly depending on the crack locations and orientations \cite{Belytschko:05,Oliver:07}.  From this point of view, discontinuous models such as conventional interface element \cite{XuX:01,Areias:01,Areias:02,Areias:05,Schrefler:02}, Strong Discontinuity embedded Approach (SDA or E-FEM) and localization \cite{Oliver:04,Armero:02,Cazes:01,Radulovic:01,Oliver:11,Riccardi:01,Motamedi:01,Cervera:04,Saloustros:01,Theiner:01}, nodal enrichment method eXtended Finite Element Method (X-FEM) \cite{Bordas:01,Moes:02,Sukumar:01,Moes:01,Meschke:01,WuJianying:01,Holzapfel:01,Holl:01} and phantom node method \cite{HansboHansbo,SongandBelytschko,Chau-Dinh2012242} could be more suitable for especially simulating the damage process of quasi-brittle materials, which holds much less mesh dependence than continuous method.  Correspondingly, when using discontinuous models for self-healing quasi-brittle materials, as a pseudo reverse damage process, healing should also be considered in a discontinuous form e.g. in the form of traction-separation law \cite{Barenblatt1962} for describing the relationship between crack opening and traction inside the cracks instead of strain and stress. 

In this paper, we present a softening-healing law in the form of traction-separation which is naturally compatible with discontinuous numerical models for quasi-brittle materials.  The law is built based on simple assumptions and only a few healing parameters with physical background are introduced.  The factors of trigger and maturity of the healing process, despite of other forms of stimuli in practices such as optical \cite{Ghosh:01} and electrical \cite{Kowalski:01}, are assumed to be crack opening and time respectively.  An SDA presented by us before in \cite{Yiming:11} is taken as the numerical framework for testing the softening-healing law, which of course does not restrict the implementation of the law in other numerical models.  Through back analysis regarding experimental results of a loading-reloading test of a self-healing concrete beam, the range of the healing parameters are determined.  Structure such as concrete member and gravity dam model are simulated for showing the advantage by using self-healing materials in engineering practices.

The remaining parts of the paper is organized as in Section~\ref{sec:model}, the proposed softening-healing law is presented, by assuming the traction-separation law of self-healing material depends on the properties of the healing agent as well as the original quasi-brittle material, then the formulation of SDA in our published literature as the numerical framework is briefly introduced.  In Section~\ref{sec:na}, the range of the material parameters of the self-healing material are determined by back analysis of loading-reloading bending tests of self-healing concrete beams.  Then, a tension-shear test is numerically analyzed for evaluating the healing efficiency of healing agent at different loading stage.  Finally a gravity dam model is simulated for showing the different working behaviors of the engineering structures built with self-healing and non-healing materials.  This paper closes with concluding remarks given in Section~\ref{sec:CC}.

\section{Model}
\label{sec:model}
\subsection{A softening-healing law}
The relationship of the crack opening and the traction between two surface of the crack is built based on a mixed-mode traction-separation \cite{Camacho:01,Meschke:01} law.  In 2D conditions, the crack opening is defined as 
\begin{equation}
\zeta=\sqrt{\zeta_n^2+\beta^2\zeta_t^2},
\label{eq:zetaeq}
\end{equation}
with $\beta$ governing the contribution of Mode I and Mode II crack opening.  According to \cite{Mariani:01,Belytschko:02}, $\beta=1$ is well-suited for quasi-brittle materials such as concrete, which will be considered throughout this paper.  $\zeta_n$ and $\zeta_t$ are the crack opening or crack width along normal and parallel directions of the crack path, with corresponding unit vector denoted as $\mathbf{n}$ and $\mathbf{t}$.

\begin{figure}[htbp]
	\centering
	\includegraphics[height=0.4\textwidth]{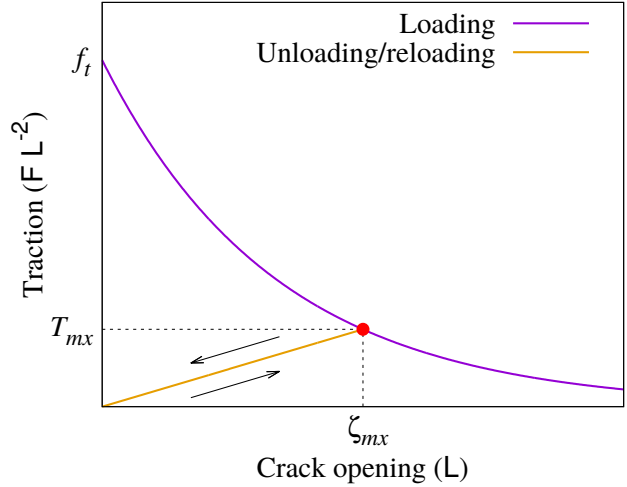}
	\caption{Original traction-seperation law}
	\label{fig:TScurve}
\end{figure}

Meanwhile, the traction components along $\mathbf{n}$ and $\mathbf{t}$, $T_n$ and $T_t$ are obtained as
\begin{equation}
\begin{aligned}
&T_n=T\frac{\zeta_n}{\zeta},~ T_t=T\frac{\zeta_t}{\zeta}\\
&\mbox{with }\\
&T\left(\zeta \right)=\left\{\begin{array}{ll}
TL\left(\zeta \right)=f_t \ \mbox{exp}\left(-\cfrac{f_t}{G_f}\zeta\right),& \mbox{loading},\\
TU\left(\zeta \right)=\cfrac{T_{mx}}{\zeta_{mx}}\zeta,& \mbox{unloading/reloading},
\end{array}\right.		
\end{aligned}
\label{eq:Traction}
\end{equation}
where $f_t$ is the uniaxial tensile strength and $G_f$ is the fracture energy, $\zeta_{mx}$ is the maximum crack opening the crack ever experienced, and $T_{mx}=TL\left(\zeta_{mx}\right) $ is the corresponding traction, as illustrated in Figure~\ref{fig:TScurve}.  

Regarding a crack which firstly experienced a big crack opening and became almost traction free i.e. $T_{mx}\approx0$ then was healed by healing agent, when this crack is reopened, the new traction denoted as $H$ is attributed to the healing agent, which depends on $\zeta$ and resting time for healing $\Delta t=t-t_{r}$, with $t_{r}$ being the time when healing agent released.  If $\Delta t$ is big enough that the healing agent become completely mature, by mimicking the traction-separation law of the original quasi-brittle material, the new traction-separation law of the healed material is obtained as:

\begin{equation}
H_\infty\left(\zeta \right)=\left\{\begin{array}{ll}
HL_\infty\left(\zeta \right)=f_{h,\infty} \ \mbox{exp}\left(-\cfrac{f_{h,\infty}}{G_{h,\infty}}\zeta\right),& \mbox{loading of healed material},\\
HU_\infty\left(\zeta \right)=\cfrac{H_{hx}}{\zeta_{hx}}\zeta,& \mbox{unloading/reloading of healed material},
\end{array}\right.	
\label{eq:HealingTraction}
\end{equation}
where $f_{h,\infty}$ is the ultimate strength and $G_{h,\infty}$ is the ultimate fracture energy of the healed material which was completely damaged before healed.  $\zeta_{hx}$ is the maximum crack opening which the crack ever experienced after healing with $\zeta_{hx}\leq \zeta_{mx}$, and $H_{hx}=HL_\infty\left(\zeta_{hx}\right) $ is the corresponding traction of healed material.  Here, it is worth to emphasize that $f_{h,\infty}$ and $G_{h,\infty}$ depend not only on the material properties of the healing agent but also on the binding properties between the healing agent and the original material.  Of course, we would like to mention here that choosing exponential relation in Equations~\ref{eq:Traction} and~\ref{eq:HealingTraction} is only an option, bilinear and hyperbolic curves are also available \cite{Morel:01,Guinea:01}, which could be more appropriate for specific materials or healing agents.

Considering the time-dependent healing behaviour that the strength of healing agent depends on the mature level of healing agent i.e. healing degree $R$ as 
\begin{equation}
H\left(\zeta,\Delta t\right)=R\left(\Delta t \right)\ H_\infty\left(\zeta  \right).
\label{eq:HealingEvolution}
\end{equation}
When assuming the solidification of healing agent is a type of chemical reaction similar to the hydration process of cementitious materials \cite{Ulm:03,Ulm:01,Yiming:01,Lackner:01}, an Arrhenius-type law is appropriate for describing $\dot{R}$, indicating $\dot{R}$ depends on temperature and is proportional to the chemical affinity $\tilde{A}$ of healing agent.  Furthermore, linear increasing of $f_{h}\left(\Delta t \right)$ and $G_{h}\left(\Delta t \right)$ with healing degree is also assumed, similar to the hydration process of early-age concrete \cite{Schutter:01,BernhardPichler:01}.  Because the experimental results of $\tilde{A}$ of cementitious materials are commonly fitted with exponential function, in isothermal condition following function for $R\left(\Delta t \right)$ is assumed
\begin{equation}
R\left(\Delta t \right)=1-\mbox{exp}\left(-A_h\ \Delta t \right)=\frac{f_{h}\left(\Delta t \right)}{f_{h,\infty}}=\frac{G_{h}\left(\Delta t \right)}{G_{h,\infty}},
\label{eq:degree}
\end{equation}
in which $A_h$ is the healing speed coefficient.

\begin{figure}[htbp]
	\centering
	\includegraphics[height=0.4\textwidth]{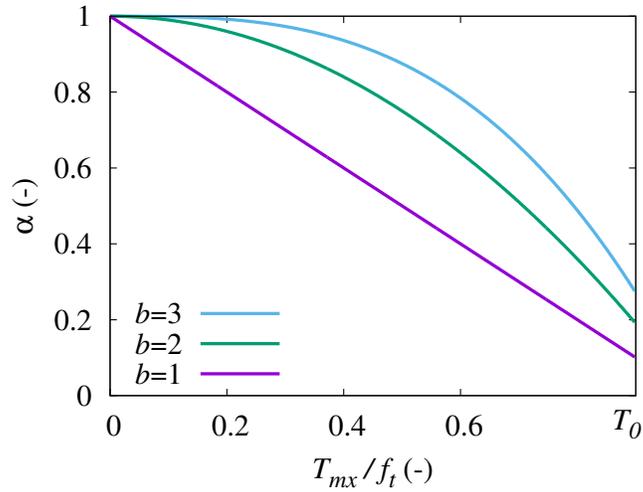}
	\caption{change of $\alpha$ with $T_{mx,r}\ / \ f_t$ regarding different values of $b$}
	\label{fig:alpha}
\end{figure}

Herein, the traction-separation law of healing agent within traction free crack is proposed.  For the cases with $T_{mx}>0$ which cannot be ignored, the system is considered as a parallel springs system that the equivalent traction $T_{eq}$ is obtained as 
\begin{equation}
\begin{aligned}
&T_n=T_{eq}\frac{\zeta_n}{\zeta},~ T_t=T_{eq}\frac{\zeta_t}{\zeta}\\
&\mbox{with }\\
&T_{eq}\left(\zeta, \Delta t\right)=T\left(\zeta\right)+\alpha\ H\left(\zeta,\Delta t\right)	
\end{aligned}
\label{eq:Tractioneq}
\end{equation}
where $\alpha$ is introduced for accounting the level of penetration and contact between the healing agent and the original material inside cohesive crack.  It is obvious that $\alpha$ is 0 when $\zeta_{mx}=0$ and will become 1 when $T_{mx}\approx0$, leading into the following assumed function
\begin{equation}
	\left\{\begin{array}{ll}
		\alpha=0, &T_{mx}> T_{0}\\
		\alpha=1-\left(T_{mx,r}\ / \ f_t\right)^b, &T_{mx}\leq T_{0}
	\end{array}\right.	
	\label{eq:alphahealing}
\end{equation}
in which, $T_{0}$ is a threshold value of traction indicating the break of carrier (glass tubes or capsules) of healing agents followed by the release and penetration of healing agent into the cohesive crack.  $T_{mx,r}$ is the value of $T_{mx}$ at $t_r$, as the crack opening when the healing agent was released.  $b$ is a coefficient indicating the contact between healing agent and the original material, with a higher value of $b$ indicating better contact, see Figure~\ref{fig:alpha}.  It can be found that $T_{0}$ indicates the sensitivity of the self-healing capsule to sense damage.  If $T_{0}$ is a big value (good sensitivity) when $b$ is a small value (bad contact and penetration), then the healing agent is released but will not perform very well since the crack opening could be a small value.  Hence, it is very important to firstly increase the value of $b$ of the healing agent in engineering practices.

 %In engineering practices, when the value of $f_t\ / \ G_f$ of quasi-brittle material is big, indicating a fast drop of softening curve, $T_{0}$ could be a very small value even when the corresponding crack opening for releasing the healing agent is not very big.  

\begin{figure}[htbp]
	\centering
	\includegraphics[height=0.4\textwidth]{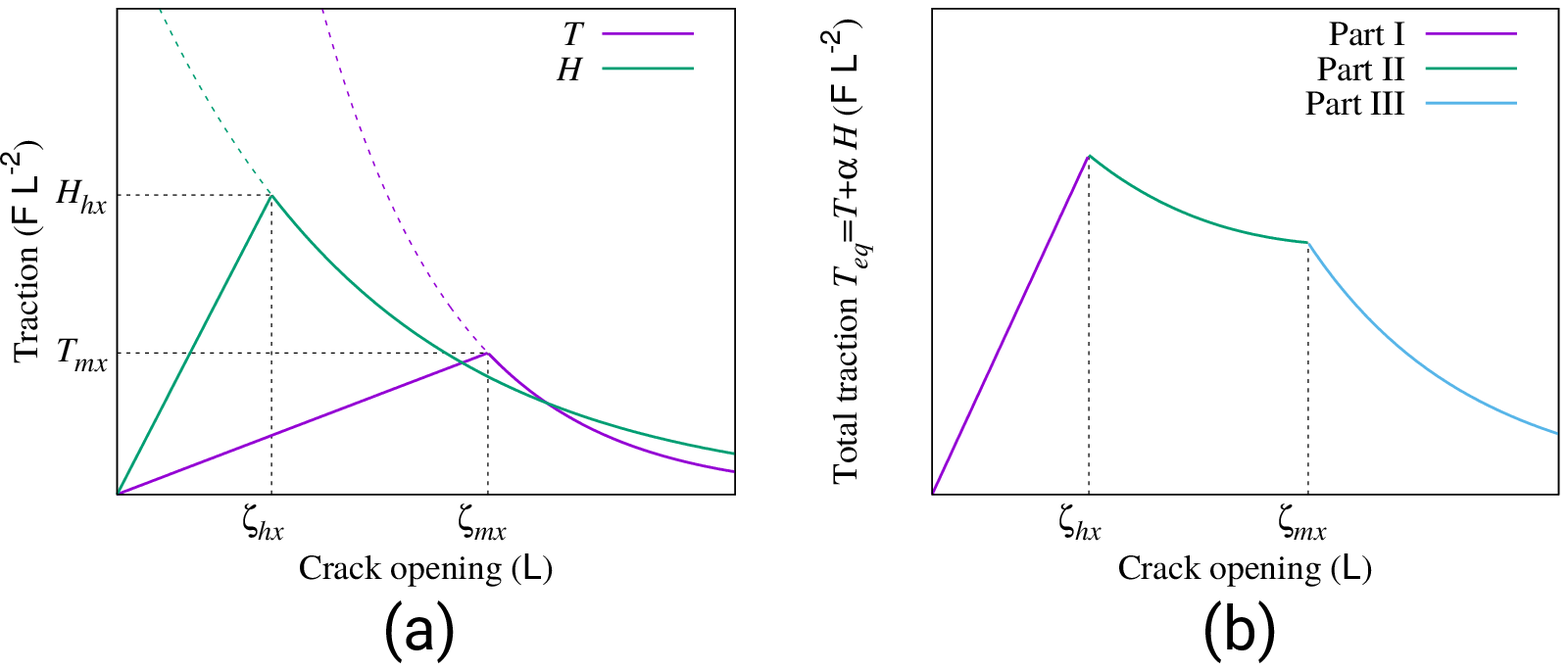}
	\caption{(a) shapes of $T\left({\zeta}\right)$, $H\left({\zeta}\right)$, (b) shape of $T_{eq}\left({\zeta}\right)$}
	\label{fig:THTEQ}
\end{figure}

Figure~\ref{fig:THTEQ} illustrates the shapes of $T\left({\zeta}\right)$, $H\left({\zeta}\right)$ and $T_{eq}\left({\zeta}\right)$, showing three parts in $T_{eq}$ 
\begin{itemize}
	\item 
    Part I: Both healing agent and original material are under unloading/reloading condition,
    \item 
    Part II: Healing agent is under loading while original material is under reloading,
    \item 
    Part III: Both healing agent and original material are under loading.
\end{itemize} 
For better understanding, the new material parameters referring to healing are listed again in Table~\ref{tab:newmaterial}.  It can be seen that the presented softening-healing law only introduce limited material parameters, which could be determined through experimental investigations.  Besides, since only the material law is changed, the new law could be very easily implemented into the current numerical models for damage analysis of quasi-brittle materials.  In the numerical studies provided in Section~\ref{sec:na}, we will focus on $f_{h,\infty}$ and $G_{h,\infty}$ the range of which are determined with back analysis of experimental results, when the other parameters are mainly assumed.

\begin{table}[htbp]
	\begin{center}
		\caption{New material parameters referring to healing}
		\label{tab:newmaterial}
		\begin{tabular}{p{1.5cm}p{1.5cm}p{3cm}p{7cm}}
			\hline
			Symbol&Unit&Presented in&For denoting\\
			\hline
			$f_{h,\infty}$&[MPa]&Equation~\ref{eq:HealingTraction}&The ultimate tensile strength of mature healing material\\
			$G_{h,\infty}$&[N m$^{-1}$]&Equation~\ref{eq:HealingTraction}&The ultimate fracture energy of mature healed material\\
			$A_h$&[h$^{-1}$]&Equation~\ref{eq:degree}&The healing speed coefficient of healing agent\\
			$T_0$&[MPa]&Equation~\ref{eq:alphahealing}&The threshold value of traction triggering the release of healing agent\\
			$b$&[--]&Equation~\ref{eq:alphahealing}&The coefficient indicating the penetration and contact property of the healing agent in the cohesive crack\\
			\hline
		\end{tabular}
	\end{center}
\end{table}

%\subsection{Healing during loading and re-healing after loading of healing agent}
%The presented softening-healing law only considers one complete loading-unloading-reloading loop.  In practices, more complicated conditions would be encountered, referring healing during loading and release of fresh healing agent and re-healing, requiring extra assumptions especially the rules for the updates of $t_r$ and $\zeta_{hx}$.  Hereby, we introduce ``threshold speed for resting of healing agent" $S_0$ the ``threshold healing degree for solidification of healing agent" $R_0$ in this section.

%Regarding $t_r$, only when the original material is under loading, healing agent will be further released and $t_r$ could be updated.  However, once the resting time is long enough for the healing degree $R(\Delta t)$ reaching $R_0$, the healing agent released at last time step is considered to be solidified and $t_r$ will not be updated anymore.  For example, there are two time step ``$t_{,i}$" and ``$t_{,i+1}$" with corresponding crack opening $\zeta_{,i}$ and $\zeta_{,i+1}$, $\zeta_{,i+1}>\zeta_{,i}$ and traction of original material $T_{,i}$ and $T_{,i+1}$, $T_{,i+1}<T_{,i}\leq T_0$.  The criterion is provided as follows

\subsection{Strong Discontinuity embedded Approach (SDA)}
The Strong Discontinuity embedded Approach (SDA) adopted in this paper is built based on standard Statically Optimal Symmetric (SOS) formulation \cite{Belytschko:01, Larsson:01, Larsson:02}.  Recently in \cite{Yiming:11}, we have proven when implemented into quadrilateral 8 node element with quadratic interpolation of the displacement field (Q8), the presented SDA model shows no stress locking as well as almost no mesh bias, which is also simple for coding and holds good numerical stability regarding standard Galerkin method.  The SDA model is briefly described as follows, when more detailed descriptions were given in \cite{Yiming:11,Feist:01,Feist:03,Mosler:01}.

\begin{figure}[htbp]
	\centering
	\includegraphics[width=0.8\textwidth]{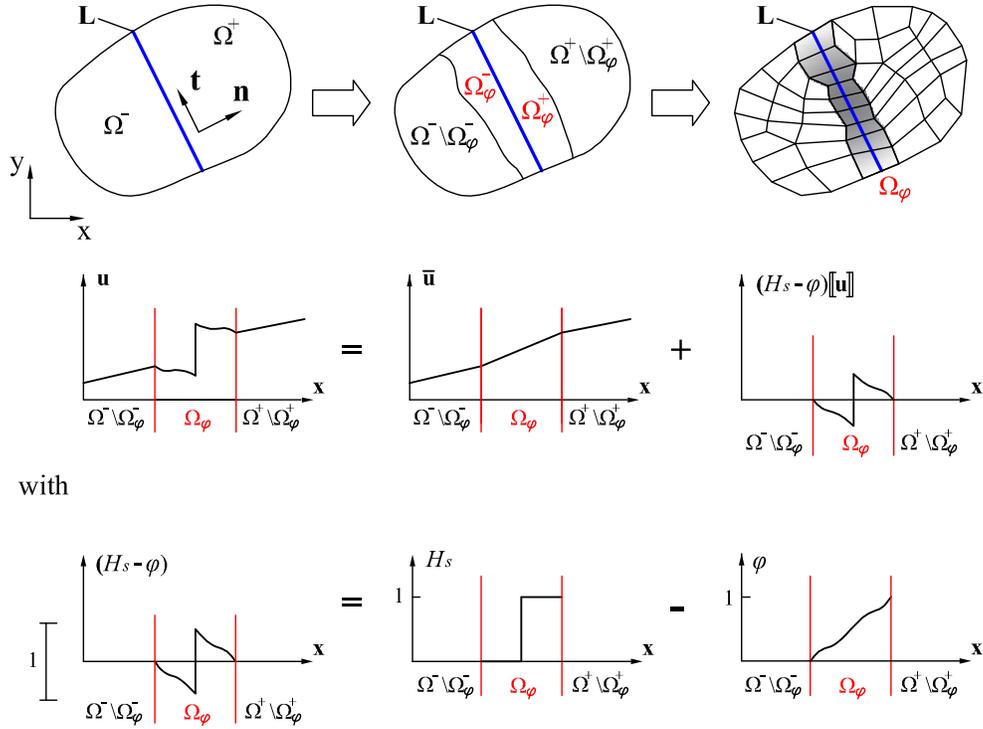}
	\caption{Domain $\Omega$ and its displacement field}
	\label{fig:Domain}
\end{figure}

Regarding a 2D domain $\Omega$ separated by a discontinuity $\mathbf{L}$ into $\Omega^+$ and $\Omega^-$ with normal and parallel unit vector $\mathbf{n}$ and $\mathbf{t}$, a localized subdomain $\Omega_\varphi$ is introduced for avoiding singularity, leading to the displacement field of $\Omega$ 
\begin{equation}
\mathbf{u}(\mathbf{x}) = \bar{\mathbf{u}}(\mathbf{x}) + [H_s(\mathbf{x})-\varphi(\mathbf{x})] \llbracket u \rrbracket,
\label{eq:dispu}
\end{equation}
where $\bar{\mathbf{u}}(\mathbf{x})$ is the regular part, $H_s(\mathbf{x})$ is the Heaviside function, and $\varphi(\mathbf{x})$ is a smooth derivable function with $\varphi(\mathbf{x})\in \left[0,1\right]$, see Figure~\ref{fig:Domain}.

The strain field \cite{Armero:01} is correspondingly obtained as 

\begin{equation}
\begin{array}{cccc}
\boldsymbol{\varepsilon}(\mathbf{x})= \nabla^S \mathbf{u}(\mathbf{x}) =&\underbrace{\nabla^S \bar{\mathbf{u}}(\mathbf{x}) - (\llbracket u \rrbracket(\mathbf{x}) \otimes \nabla \varphi)^S}&  +& \underbrace{\delta_\mathbf{L} (\llbracket u \rrbracket(\mathbf{x}) \otimes \mathbf{n})^S }, \\
&\bar{\boldsymbol{\varepsilon}}(\mathbf{x}), \forall x \in \Omega \setminus \mathbf{L}&&{\boldsymbol{\varepsilon}}_\delta(\mathbf{x}), \forall x \in \mathbf{L}
\end{array}
\label{eq:dispE}
\end{equation}
where $(\cdot)^S$ denotes the symmetric part of the tensor and $\delta_\mathbf{L}$ stands for the Dirac-delta distribution.  After assuming $\nabla\llbracket u \rrbracket(\mathbf{x})=0$ and further decomposing $\llbracket u \rrbracket(\mathbf{x})=\zeta_n(\mathbf{x}) \mathbf{n}+\zeta_t(\mathbf{x}) \mathbf{t}$, compatible strains and enhanced strains \cite{Simo:02} are obtained as

\begin{equation}
\begin{array}{cccc}
\bar{\boldsymbol{\varepsilon}}(\mathbf{x})=&\underbrace{\nabla^S \bar{\mathbf{u}}(\mathbf{x})}&-&\underbrace{\left[(\mathbf{n} \otimes \nabla \varphi)^S \zeta_n(\mathbf{x})+(\mathbf{t} \otimes \nabla \varphi)^S \zeta_t(\mathbf{x})\right]},\\
&\mbox{total strain } \boldsymbol{\varepsilon}^t&&\mbox{enhanced strain } \widetilde{\boldsymbol{\varepsilon}}
\end{array}
\label{eq:EAS}
\end{equation}

\begin{figure}[htbp]
	\centering
	\includegraphics[width=0.6\textwidth]{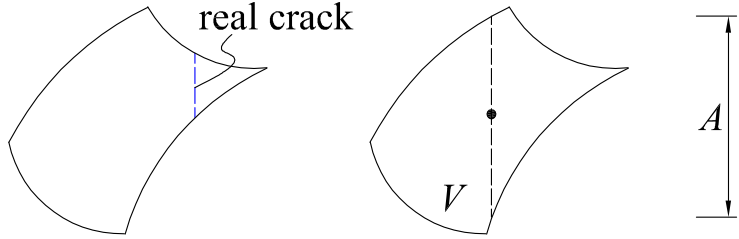}
	\caption{Determinating the effective cracking area $A$}
	\label{fig:LC}
\end{figure}

Based on SOS formulation and the conservation of dissipated energy in localized element and using reduced integration scheme, in \cite{Yiming:11} we obtained $\nabla \varphi=\mathbf{n}\ /\ l_c$ and $l_c=V\ /\ A$, where $V$ is the volume of the element and $A$ is the area of a parallel crack passing through center point, see Figure~\ref{fig:LC}, and $l_c$ corresponds to the classical characteristic length \cite{Oliver:02}, which only depends on the shape of the local element.  Once the crack direction is determined by the so called crack tracking strategy \cite{Cervera:03,Dumstorff:01,Unger:01,Oliver:03,Stolarska:01,Annavarapu:01}, $l_c$ is directly obtained.  Hence, in the local element ``$e$", following relationship is obtained

\begin{equation}
\bar{\boldsymbol{\varepsilon}}^{(e)}(\mathbf{x})\approx\sum^{n_e}_{i=1}\left(\nabla N^{(e)}_i \otimes \mathbf{u}_i\right)^S-\frac{1}{\ l_c \ }\left[(\mathbf{n} \otimes  {\mathbf{n}}) \zeta_n^{(e)}+(\mathbf{n} \otimes  {\mathbf{t}})^S \zeta^{(e)}_t\right],
\label{eq:Strain_e}
\end{equation}
where ${n_e}$ is the number of nodes of element "$e$", $N^{(e)}$ is the shape function, and $\mathbf{u}_i$ is the displacement vector of node "$i$".

\subsection{Algorithmic formulation}
Regarding the states at steps ``$i-1$" and ``$i$", extra subscript is used for showing different parameters in these two steps, for example $\zeta_i$ is the crack opening at step $i$.  Hereby, the states at ``$i-1$" are all known when most states at ``$i$" are unknown.  At step ``$i$", the trial stress state is obtained as 
\begin{equation}
\boldsymbol{\sigma}^{tr}_i=\mathbb{C}:\left\{\boldsymbol{\varepsilon}^t_i-\frac{1}{\ l_c \ }\left[(\mathbf{n} \otimes \mathbf{n}) \zeta_{n,i-1}+(\mathbf{n} \otimes  \mathbf{t})^S \zeta_{t,i-1}\right]\right\},
\label{eq:sigmatr}
\end{equation}
where $\mathbb{C}$ is the elasticity tensor.  Regarding the balance relation, the unknown incremental state variables $\Delta\zeta_{n,i}=\zeta_{n,i}-\zeta_{n,i-1}$ and $\Delta\zeta_{t,i}=\zeta_{t,i}-\zeta_{t,i-1}$ are determined from 
\begin{equation}
\left[
\begin{array}{c}
\mathbf{V}_1\\
\mathbf{V}_2
\end{array}
\right]:\boldsymbol{\sigma}^{tr}_i-\frac{1}{\ l_c \ }
\left[
\begin{array}{cc}
\mathbf{V}_1:\mathbb{C}:\mathbf{V}_1&\mathbf{V}_1:\mathbb{C}:\mathbf{V}_2\\
\mathbf{V}_2:\mathbb{C}:\mathbf{V}_1&\mathbf{V}_2:\mathbb{C}:\mathbf{V}_2
\end{array}
\right]
\left[
\begin{array}{c}
\Delta \zeta_{n,i}\\
\Delta \zeta_{t,i}
\end{array}
\right]
-\left[
\begin{array}{c}
T_n\\
T_t
\end{array}
\right]=\mathbf{0}
\label{eq:yielditeration}
\end{equation}
with $\mathbf{V}_1=(\mathbf{n}\otimes\mathbf{n})$ and $\mathbf{V}_2=(\mathbf{n}\otimes\mathbf{t})^S$.

Regarding infinitesimal multisurface plasticity \cite{Simo:05,Mosler:04,Mosler:07}, the corresponding elastoplastic tangent moduli $\mathbb{C}^{ep}$ at step $i$ is  
\begin{equation}
\mathbb{C}^{ep}=\frac{d \boldsymbol{\sigma}_i}{d \boldsymbol{\varepsilon}^t_i}=\mathbb{C}-\mathbb{S},
\label{eq:tangentModuli}
\end{equation}
where
\begin{equation}
\mathbb{S}=\mathbb{C}:
\left[\begin{array}{cc}
\mathbf{V}_1&\mathbf{V}_2\\
\end{array}\right]:
\left(\mathbf{G}+l_c\ \mathbf{D}\right)^{-1}:
\left[\begin{array}{c}
\mathbf{V}_1\\
\mathbf{V}_2\end{array}
\right]:
\mathbb{C}
\label{eq:tangentModuliS}
\end{equation}
with
\begin{equation}
\mathbf{G}=\left[
\begin{array}{cc}
\mathbf{V}_1:\mathbb{C}:\mathbf{V}_1&\mathbf{V}_1:\mathbb{C}:\mathbf{V}_2\\
\mathbf{V}_2:\mathbb{C}:\mathbf{V}_1&\mathbf{V}_2:\mathbb{C}:\mathbf{V}_2
\end{array}
\right], \nonumber
\label{eq:tangentModuli2}
\end{equation}
and
\begin{equation}
\mathbf{D}=\mathbf{D}_o+\alpha \ \mathbf{D}_h
\label{eq:tangent2}
\end{equation}
where
\begin{equation}
\begin{array}{l}
\mathbf{D}_o=-\frac{T}{\zeta}
\left[\begin{array}{cc}
\frac{\zeta_n^2}{\zeta^2}+\frac{f_t\zeta_n^2}{G_f\zeta}-1&
\frac{\zeta_n\zeta_t}{\zeta^2}+\frac{f_t\zeta_n\zeta_t}{G_f\zeta}\\
\frac{\zeta_n\zeta_t}{\zeta^2}+\frac{f_t\zeta_n\zeta_t}{G_f\zeta}&
\frac{\zeta_t^2}{\zeta^2}+\frac{f_t\zeta_t^2}{G_f\zeta}-1\\
\end{array}\right] \mbox{  for loading of original material},   \\
\mbox{and}\\
\mathbf{D}_o=
\frac{T_{mx}}{\zeta_{mx}}
\left[\begin{array}{cc}
1&0\\
0&1
\end{array}\right] \mbox{  for unloading/reloading of original material},  
\end{array}
\label{eq:tangent3}
\end{equation}
and
\begin{equation}
\begin{array}{l}
\mathbf{D}_h=-\frac{H}{\zeta}
\left[\begin{array}{cc}
\frac{\zeta_n^2}{\zeta^2}+\frac{f_{h,\infty}\zeta_n^2}{G_{h,\infty}\zeta}-1&
\frac{\zeta_n\zeta_t}{\zeta^2}+\frac{f_{h,\infty}\zeta_n\zeta_t}{G_{h,\infty}\zeta}\\
\frac{\zeta_n\zeta_t}{\zeta^2}+\frac{f_{h,\infty}\zeta_n\zeta_t}{G_{h,\infty}\zeta}&
\frac{\zeta_t^2}{\zeta^2}+\frac{f_{h,\infty}\zeta_t^2}{G_{h,\infty}\zeta}-1\\
\end{array}\right] \mbox{  for loading of healing agent},   \\
\mbox{and}\\
\mathbf{D}_h=
\frac{H_{hx}}{\zeta_{hx}}
\left[\begin{array}{cc}
1&0\\
0&1
\end{array}\right] \mbox{  for unloading/reloading of healing agent}.
\end{array}
\label{eq:tangent4}
\end{equation}

\section{Numerical analysis}
\label{sec:na}
\subsection{Bending test for back analysis}
\begin{figure}[htbp]
	\centering
	\includegraphics[width=0.8\textwidth]{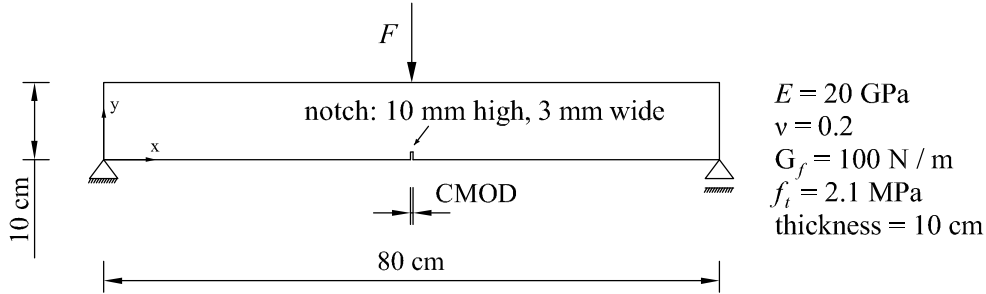}
	\caption{Set-up of the bending test and material properties of concrete}
	\label{fig:BendingModel}
\end{figure}

For determining the material parameters of healing agent, the results of three point bending test of a self-healing concrete beam provided in \cite{Tsangouri:01} are taken.  The set-up of the test and material parameters of concrete is shown in Figure~\ref{fig:BendingModel} \cite{Tsangouri:01,Tsangouri:02}, with a mixture ratio of concrete as water/cement/sand/gravel = 1/2/4.47/8.53.  The material properties of the concrete were not directly provided in \cite{Tsangouri:01}, but taken by us based on other fracture tests with similar concrete mixture in \cite{Rots:01,Winkler:01}.  Glass capsules filled with two-component expansive polyurethane-based healing agent are casted inside the concrete beam.  The simple supported beam was vertically loaded in the middle with a displacement speed 0.04 mm/min, considered as static loading condition.  The force $F$ and Crack Mouth Opening Displacement (CMOD) of the notch were recorded.  The beam was firstly loaded until CMOD reached 0.3 mm, which was completely unloaded and stored for 24 hours.  Thereafter the loading procedure was repeated for assessing the performance of healing.  Considering the loading condition, $t_r$ is taken as the time when CMOD=0.3 mm was firstly reached.

\begin{figure}[htbp]
	\centering
	\includegraphics[width=0.9\textwidth]{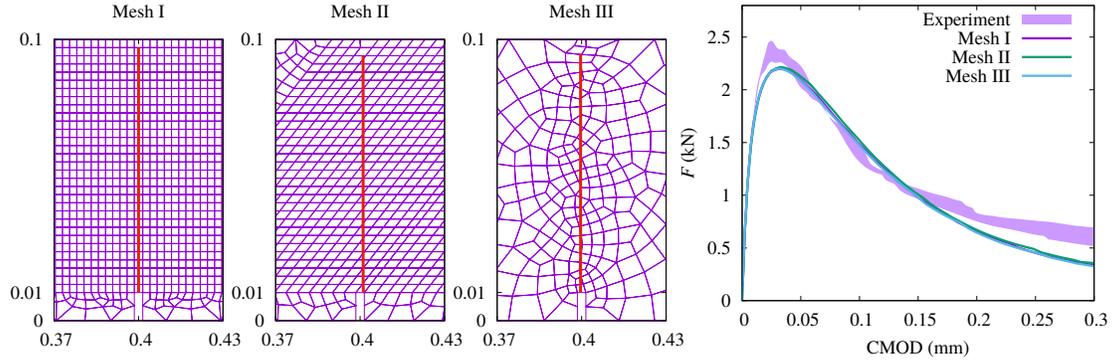}
	\caption{Force-CMOD curves of the bending test, considering different discretizations}
	\label{fig:BendingMesh}
\end{figure}

\begin{figure}[htbp]
	\centering
	\includegraphics[width=0.9\textwidth]{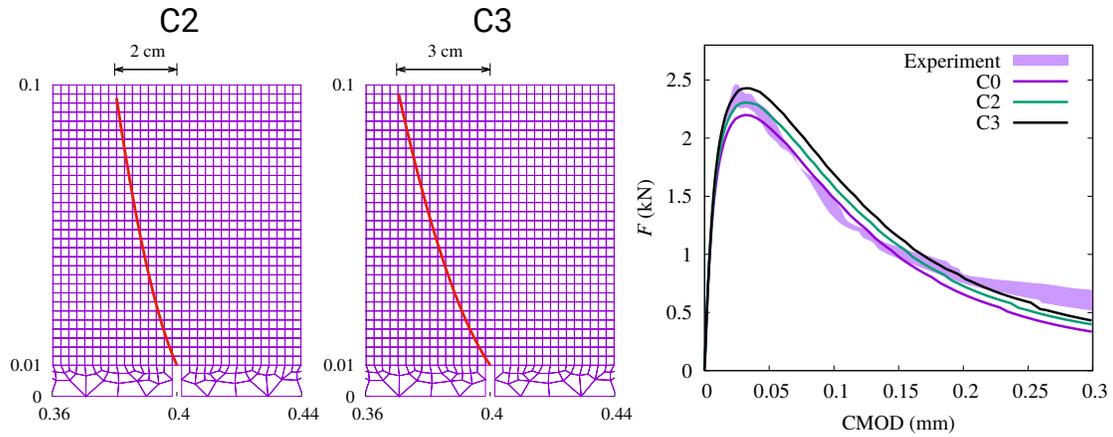}
	\caption{Force-CMOD curves of the bending test (case with straight path denoted as C0), considering curved crack paths (curve equation $\sqrt{y}+a\left(x-x_0\right)=0$)}
	\label{fig:BendingPath}
\end{figure}

In this test, the crack tracking strategy is not used.  Firstly a straight crack path is prescribed, regarding three different discretization.  The force-CMOD curves shown in Figure~\ref{fig:BendingMesh} indicate the adopted SDA model providing results with no mesh bias, as before pointed out in \cite{Yiming:11}.  Hence, in the following analysis, only Mesh I is used.  Furthermore, when in reality the crack path is commonly not a straight line as prescribed, curved crack paths described by equation $\sqrt{y}+a\left(x-x_0\right)=0$ with $a$ and $x_0$ as parameters are also taken into account, the obtained force-CMOD curves of which are shown in Figure~\ref{fig:BendingPath} (case with straight path denoted as C0).  In the back analysis of the healing-reloading test, curved paths are considered together with the straight crack path.   

\begin{figure}[htbp]
	\centering
	\includegraphics[width=0.6\textwidth]{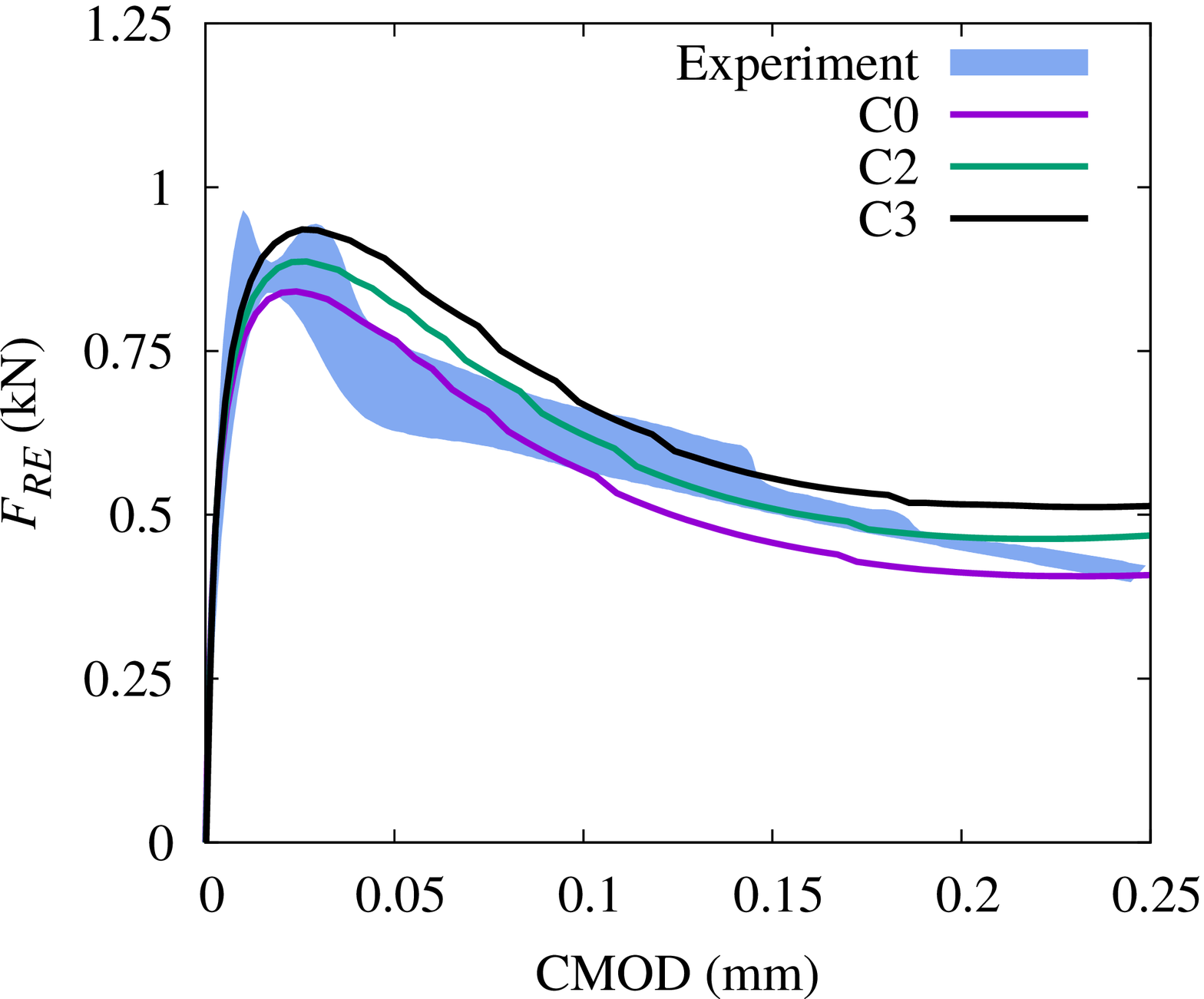}
	\caption{Force-CMOD curves of the reloading test after healing, considering straight and curved crack paths}
	\label{fig:reloading}
\end{figure}

As mentioned before, we will mainly focus on $f_{h,\infty}$ and $G_{h,\infty}$ in this section.  When assuming the healing degree of the healing agent reached 90\% after 24 hours resting, as $A_h=0.096\ \mbox{h}^{-1}$, the evolution of healing agent is shown in Figure~\ref{eq:HealingEvolution}.  Further by taking $b=2.0$ and $T_0=0.5\ f_t$, $f_{h,\infty}=0.7\ \mbox{MPa}$ and $G_{h,\infty}=42\ \mbox{N/m}$ are obtained from the back analysis.  The numerically-obtained force $F_{RE}$-CMOD curves of reloading are shown in Figure~\ref{fig:reloading}, indicating acceptable agreements between numerically and experimental-obtained results.  
  
\begin{figure}[htbp]
	\centering
	\includegraphics[width=0.55\textwidth]{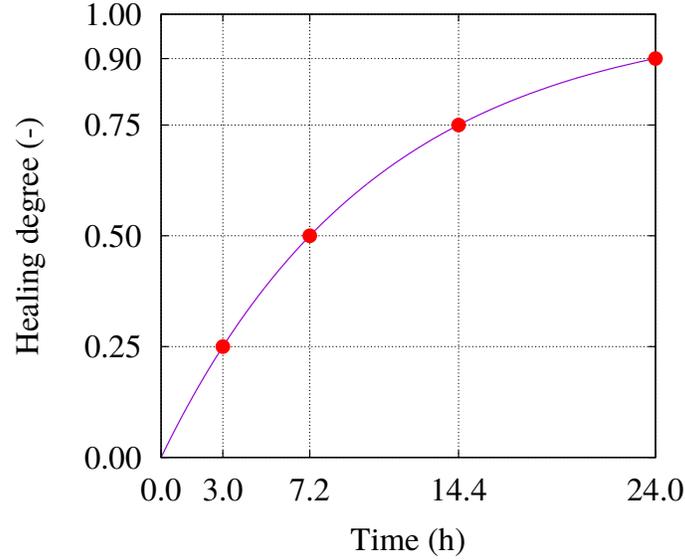}
	\caption{Assumed evolution of maturity of the healing agent based on Equation~\ref{eq:degree}}
	\label{fig:Healingdegree}
\end{figure}

\subsection{Tension-shear test}
\begin{figure}[htbp]
	\centering
	\includegraphics[width=0.5\textwidth]{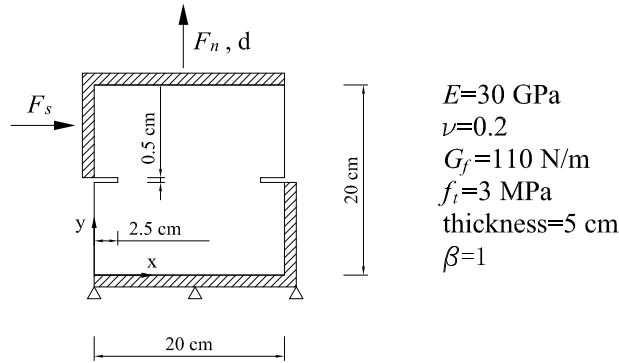}
	\caption{Geometric properties, material parameters, and support conditions of tension-shear test}
	\label{fig:MixModel}
\end{figure}

\begin{figure}[htbp]
	\centering
	\includegraphics[width=0.65\textwidth]{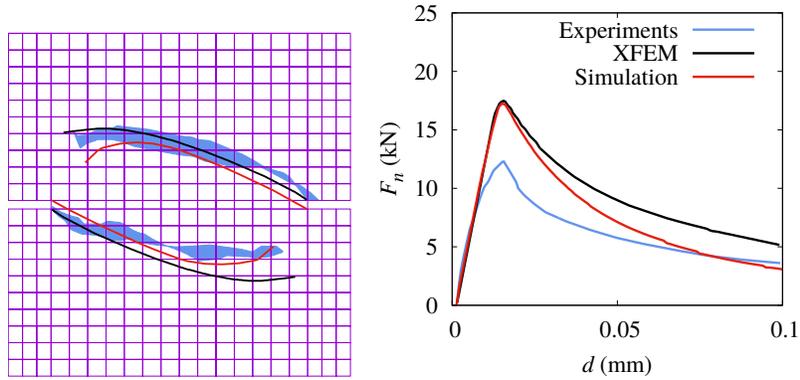}
	\caption{Original crack paths and load-displacement curves of tension-shear test, compared with experimental results \cite{NooruMohamed:01} and numerical results obtained by the XFEM \cite{Meschke:01}}
	\label{fig:MixRE}
\end{figure}

The second example is the tension-shear test \cite{NooruMohamed:01}.  The model, loading and support conditions are shown in Figure~\ref{fig:MixModel}, with the "Load Path 4" in \cite{NooruMohamed:01} and horizontal load $F_s=10$kN considered.  First, $F_s$ is applied and kept constant.  Thereafter, $F_n$ is gradually increased, leading to a vertical displacement $d$ and to cracking of the structure.  The energy-based crack tracking strategy \cite{Yiming:11} is used for predicting the crack propagations.  The original crack paths and the load-displacement curves of non-healing material are shown in Figure~\ref{fig:MixRE}, with the comparisons with experimental results \cite{NooruMohamed:01} and numerical results obtained by the XFEM \cite{Meschke:01}.

\begin{figure}[htbp]
	\centering
	\includegraphics[width=1\textwidth]{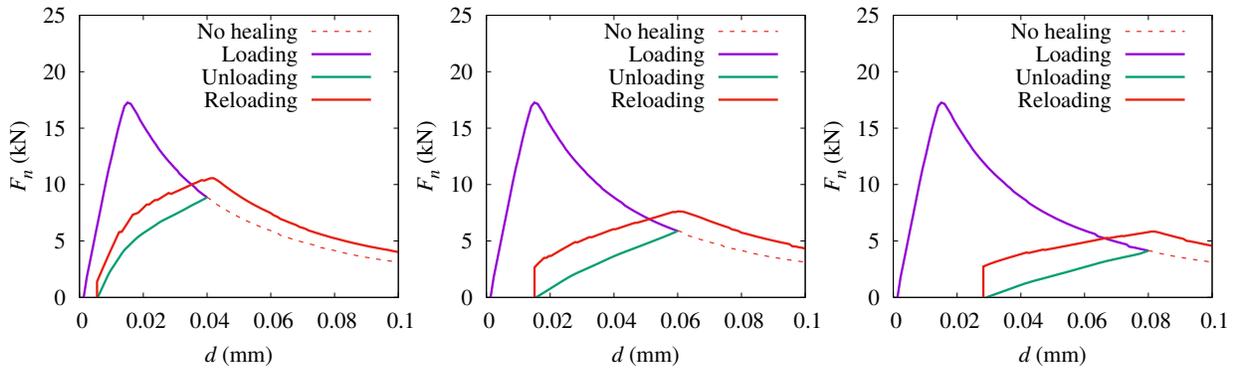}
	\caption{Load-displacement curves of tension-shear test when firstly unloaded at $d$=0.04 mm, 0.06 mm and 0.08 mm, then healed and reloaded after 24h}
	\label{fig:MixHealingRe}
\end{figure}

\begin{figure}[htbp]
	\centering
	\includegraphics[width=0.6\textwidth]{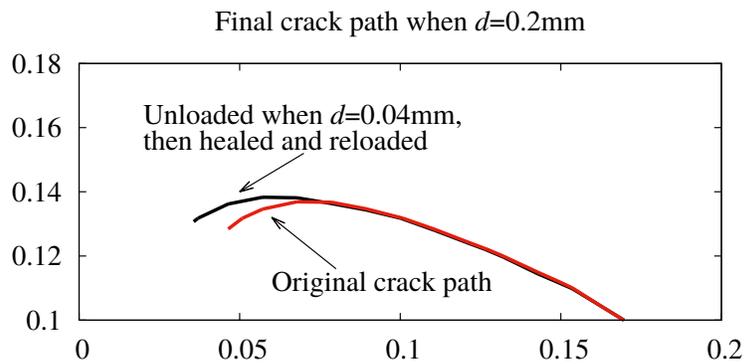}
	\caption{Influence of healing on the further propagated crack path}
	\label{fig:MixPath}
\end{figure}

Regarding the self-healing parameters, $f_{h,\infty}=0.7\ \mbox{MPa}$ and $G_{h,\infty}=50\ \mbox{N/m}$ are taken, similar to the values obtained from the back analysis.  Meanwhile, $A_h=0.096\ \mbox{h}^{-1}$, $b=2.0$ and $T_0=0.5\ f_t$ are assumed.  Three reloading cases with target $d$=0.04 mm, 0.06 mm and 0.08 mm are considered.  Firstly the structure is loaded till the targeted $d$ is reached when $t_r$ is taken as the corresponding time, then $F_n$ is reduced to zero and begins to increase again after 24 hours, in the whole procedure of which $F_s$ keeps as 10 kN.  The Load-displacement curves are shown in Figure~\ref{fig:MixHealingRe}, indicating that the recovery of mechanical strength was captured by the presented softening-healing model.  On the other hand, it could be found when using the model, the contraction is obtained even when the structure is not fully unloaded, which was not completely proven in experimental investigations.  More studies will be carried on in the future.  On the other hand, the influence of healing on the further propagated crack path is also evaluated.  Figure~\ref{fig:MixPath} shows the different crack path of the cases with and without healing, indicating the healing process slightly reduces the curvature of the crack path.  Nevertheless, we would like to mention that this difference can be almost ignored in this case.

\subsection{A gravity dam model}
\begin{figure}[htbp]
	\centering
	\includegraphics[width=0.85\textwidth]{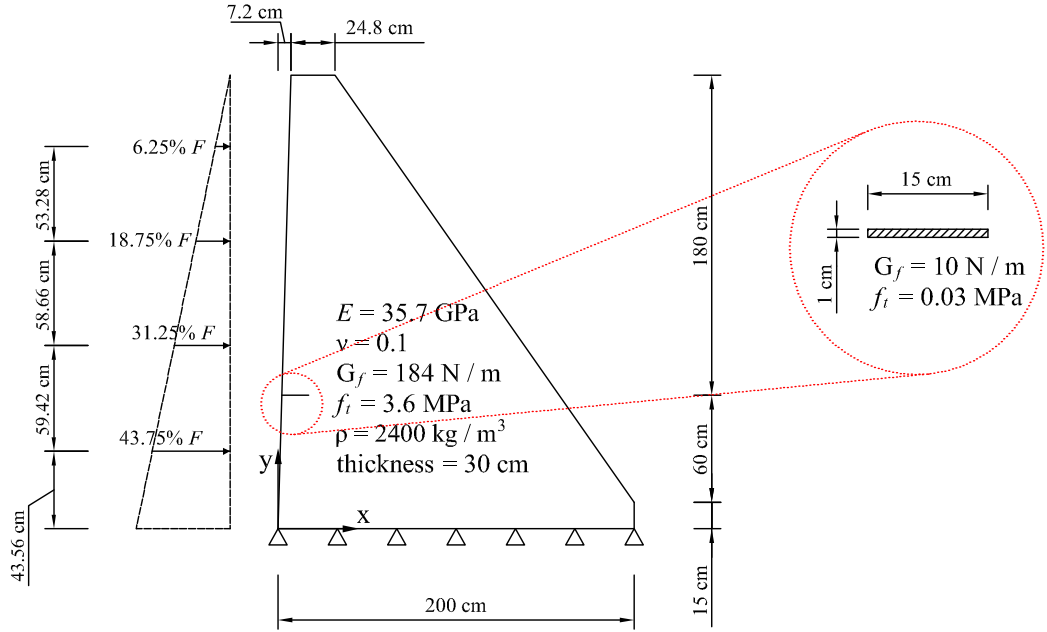}
	\caption{Geometric properties, material parameters, and support conditions of the gravity dam model}
	\label{fig:DamModel}
\end{figure}

\begin{figure}[htbp]
	\centering
	\includegraphics[width=0.9\textwidth]{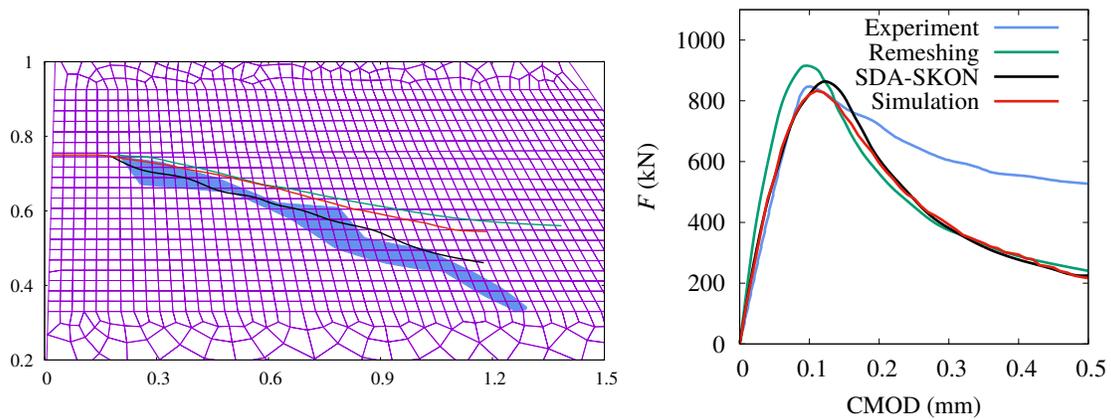}
	\caption{Original load-CMOD curves and crack paths of the dam model, compared with experimental results \cite{Barpi:01} and numerical results obtained by remeshing \cite{Areias:02} and strain injection (upgraded SDA-SKON) \cite{DiasIF:01} methods}
	\label{fig:DamREO}
\end{figure}

Hydraulic structures such as dams are very suitable for using self-healing materials, when these structures are commonly
\begin{itemize}
	\item 
	designed and constructed for over decades of long term working,
	\item
	suffered from initial imperfection such as shrinkage cracking of early-age concrete and durability problems such as leakage,
	\item
	in large scale and partially under water, which are expensive for traditional maintaining and repairing involving human effort.
\end{itemize}
Hence, we simulate the gravity dam model experimentally investigated in \cite{Barpi:01} for illustrating the advantages by using self-healing materials regarding the mechanical behavior.  The model was shown in Figure~\ref{fig:DamModel}, with the notch considered as a kind of imperfection of the original structure which could be healed by using self-healing materials.  Similar to the experimental set-up, in the numerical model, the load was simplified by concentrated forces at 4 heights.  The arc-length method is used for inducing a constant increment for the CMOD of the notch.  The obtained force-CMOD curves and crack paths of non-healing are shown in Figure~\ref{fig:DamREO}, comparing to the numerical results in \cite{Areias:02,DiasIF:01}.  

\begin{figure}[htbp]
	\centering
	\includegraphics[width=0.5\textwidth]{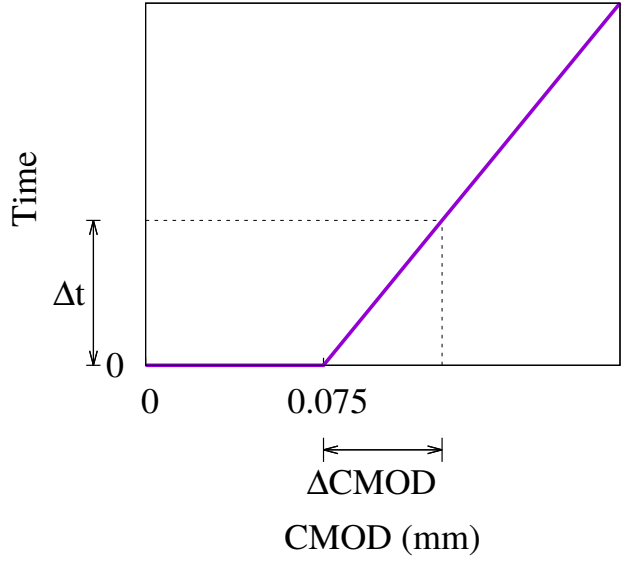}
	\caption{The loading step used in the dam model}
	\label{fig:step}
\end{figure}

\begin{figure}[htbp]
	\centering
	\includegraphics[width=0.9\textwidth]{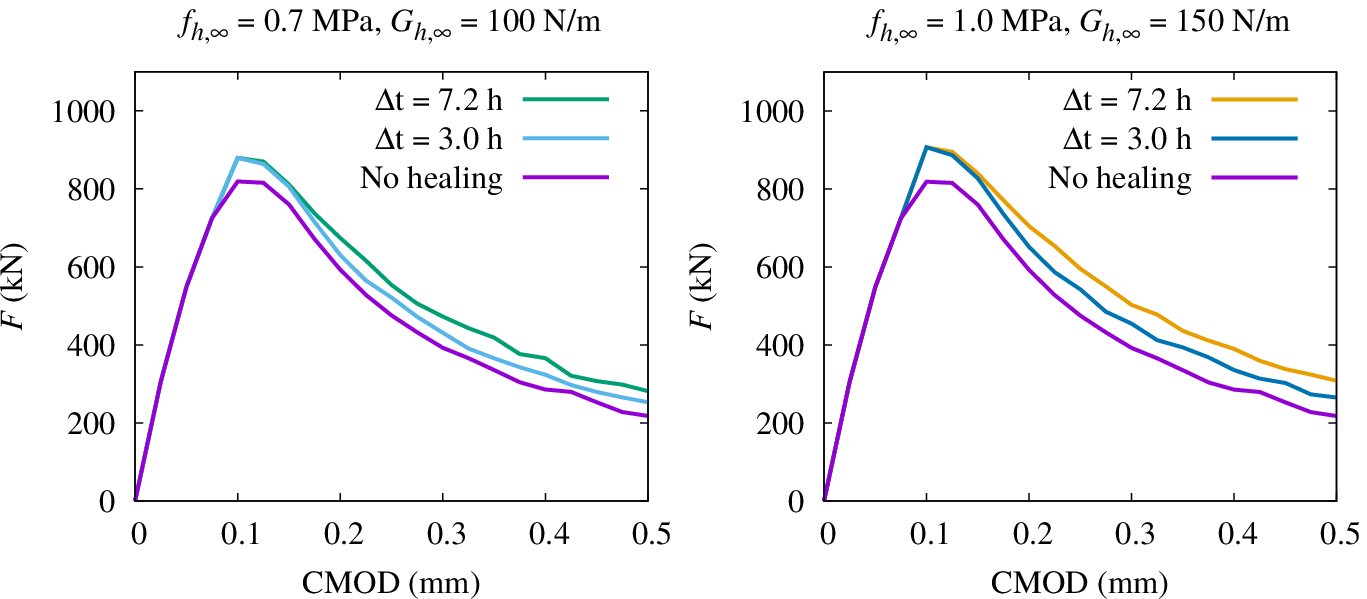}
	\caption{Load-CMOD curves of the dam model with self-healing materials}
	\label{fig:DamHeal}
\end{figure}

In the numerical analysis with self-healing material, we consider CMOD=0.075 is the initial step for loading and keep a constant increment $\Delta\mbox{CMOD}=0.025$mm.  The time for every increment is $\Delta t$, as shown in Figure~\ref{fig:step}.  Furthermore, an explicit strategy is used for obtaining $t_r$, that if the crack further propagates between calculation step $i-1$ and $i$, from position $\mathbf{x}_{i-1}$ to $\mathbf{x}_{i}$ with corresponding time $t_{i-1}$ and $t_i$ and $T_{mx}\leq T_0$, then $t_r=t_i$ is considered for the crack $\mathbf{x}_{i-1}\rightarrow \mathbf{x}_{i}$.  A sensitive healing agent with good penetration property is assumed, with $T_0=0.9\ f_t$ and $b=3.0$.  Considering $\Delta t$=3.0 h and $\Delta t$=7.2 h, the load-CMOD curves of the self-healing structures are show in Figure~\ref{fig:DamHeal}, indicating dam model with self-healing material holds better bearing capacity and post peak behavior, even when $\Delta t$=3.0 h.

\section{Conclusions}
\label{sec:CC}
In this paper, aiming at the traction-separation property of the self-healing quasi-brittle materials, we present a softening-healing law.  The law is built based on simple assumptions with limited material parameters introduced.  The new parameters take into account the properties of maturity, strength, trigger effect, contact and penetration with the original material of the healing agent.  A strong discontinuity embedded approach presented by us before is used for testing the softening-healing law.  The general range of the healing parameters are obtained by back analysis of a self-healing beam bending test.  Then more complicated examples including the tension-shear member test and the dam model test are used for showing the different load-deformation behaviors of the non-healing and self-healing materials, indicating the effectiveness of the softening-healing law as well as the applicability of the self-healing material in engineering practices.

\section*{Acknowledgement}
The authors gratefully acknowledge financial support by Alexander von Humboldt Foundation Germany, through Sofja Kovalevskaja Award, year 2015 winner Dr Xiaoying Zhuang.